\documentstyle[11pt]{article}
\begin{document}
\title{Bifermionic Superfluidity in Toroidal Optical Lattices}
\author{K.G. Petrosyan\footnote{pkaren@yerphi.am} \\
\it{Department of Theoretical Physics} \\
\it{Yerevan Physics Institute, Yerevan 375036, Armenia}}
\date{\today}
\maketitle

\begin{abstract}
We consider a gas of neutral fermions trapped in a specific
optical trap that provides a tight confinement of a Fermi gas in a
torus with a potential periodic along the azimuthal direction. The
effective model is interacting fermions moving in a periodic
potential along the ring. We show that the model demonstrates a
{\it novel type of superfluidity} different from the BCS
mechanism. This pure 1D quantum gas in a ring can also form
crystalline structures with both ferro- and antiferro-magnetic
type orders. Possibilities of realization of quantum crystals and
their application for quantum computing are discussed.
\end{abstract}


{\it Introduction}. After the realization of Bose-Einstein
condensation in trapped alkali gases \cite{bec} and recent
achievements in producing degenerate Fermi gases \cite{jin} there
is a growing interest in fundamental properties of these quantum
objects and their applications. Pure samples of fermions in
magnetic and optical traps provide with a unique possibility to
test a number of quantum statistical physics effects. That also
advances towards realizations of new quantum devices such as
quantum computers \cite{qc} and intense fermionic beam generators
\cite{fal, potting}. Especial focus since recently has been on
ultracold quantum gases confined in optical lattices. Successful
experiments with bosons \cite{optlat} and production of a Fermi
gas of atoms in an optical lattice \cite{modugno} inspired several
theoretical works related to cold atoms trapped in optical
lattices \cite{jaksch}. An atomic Bose-Fermi mixture in an optical
lattice was considered in \cite{lew}. Exciting issues of possible
control of interatomic interaction in optical lattices were
addressed in \cite{duan}. Among experimental goals in this area
most challenging have been getting a two-component Fermi gas in
the BCS state \cite{stoof} and the recently realized creation of
ultracold molecules from a Fermi gas of atoms \cite{nature}. A
possibility of a superfluid phase transition in a single-component
polarized Fermi gas of dipolar molecules was studied in
\cite{dipolar}. Further elaboration of the direction will
certainly shed more light on problems of quantum physics of {\it
composite} particles. Formation of crystalline structures from the
trapped quantum gases is yet another important direction linking
the mesoscopic physics of ultracold atoms with condensed matter
physics \cite{jaksch}. Recently a new technique was suggested for
preparation of atomic crystals in an optical lattice \cite{rabl}.
The proposed scheme would make it possible to directly engineer
states with a specific number of atoms per site in optical
lattices thus realizing atomic crystals. In the present paper we
demonstrate a novel type of superfluidity different from the BCS
mechanism and consider a possibility of obtaining atomic crystals
from samples of neutral fermions trapped in a specific optical
trap. We suggest to use an optical trap which consists of two {\it
co}-propagating lasers beams with a Laguerre-type transverse
profile. The chosen profile guarantees the Fermi gas to be trapped
in an effectively 1D ring. In addition to that the presence of two
laser beams creates a potential periodic along the ring's
azimuthal direction. So far we have a 1D two species neutral Fermi
gas trapped in a ring which experiences a periodic potential. The
system can be also seen as a two-species Fermi gas sample trapped
in a 1D cyclic optical lattice.

{\it Toroidal Optical Trap}. Let us consider the optical potential
created by two {\it co-}propagating focused laser fields which
have the same Gaussian-Laguerre transverse profile, but with {\it
opposite} circulations for the azimuthal phase $\phi$ (with
topological charges $\pm m$), {\it i.e.}
\begin{eqnarray}
\hat E_{\pm }=\hat e E_0\left(\frac{\rho}{\sqrt{2}w}\right)^{|m|}
e^{\pm im\phi}L_{p}^{|m|}\left( \frac{\rho^{2}}{2w^2}\right)
e^{-\rho^{2}/4w^{2}}e^{ikz}e^{i\omega_L t},
\end{eqnarray}
where $w$ is the width of the Gaussian prefactor for the
amplitudes, and $\rho$ is the radial coordinate in the transverse
direction. Both fields are assumed to have the same polarization
vector $\hat e$, thus driving the same atomic transitions. The
trapping potential established from these two fields takes the
following simple form
\begin{eqnarray}
V_{\Omega}(\vec{r})\propto |\hat E_{+}+\hat E_{-}|^{2} = {1\over
2}\hbar{\Omega^2(\rho)\over \Delta} \left(1+\cos \left[2m\phi
\right]\right), \label{fort}
\end{eqnarray}
with now the effective detuning $\Delta=\omega_L-\omega_0$ is the
laser detuning from an electronic transition at $\omega_0$. We
have defined the Rabi frequency,
\begin{eqnarray}
\hbar\Omega(\rho)\equiv 2 E_0 ({\hat e\cdot\vec d})
\left(\frac{\rho}{\sqrt{2}w}\right)^{|m|}
L_{p}^{|m|}\left(\frac{\rho^2}{2w^2}\right)e^{-\rho^{2}/4w^{2}}
\end{eqnarray}
with $\vec d$ the electronic dipole of the far off resonant
transition. We have conveniently chosen the light field to be
propagating in the z-direction. In a realistic setup, one may want
to use an intersecting light sheet to confine the axial motion in
the z-direction, such that an effective two dimensional motion
results. Additional optical potential terms can be introduced to
confine the system into a torus shaped one-dimensional motion
along the ring. Yet another scheme to produce a toroidal trap was
considered in \cite{wright}.

{\it Neutral Fermi Gas confined in a ring}. Let us consider a gas
of fermions confined in a trap which interact with each other by
two-body collisions ($s$-wave scattering). The Hamiltonian reads
\begin{eqnarray}
H=\sum_{\sigma=\uparrow \downarrow} \int d^3 \vec r
\left(\psi_{\sigma}^{\dag} (\vec r)
\left[-\frac{\hbar^2}{2M}\nabla^2 - \mu_{\sigma} + V_{\sigma}
(\vec r) + V_{\Omega}(\vec r) \right] \psi_{\sigma}(\vec r)
+\frac{g}{2} \psi_{\sigma}^{\dag}(\vec
r)\psi_{-\sigma}^{\dag}(\vec r) \psi_{-\sigma}(\vec
r)\psi_{\sigma}(\vec r)\right)
\end{eqnarray}
Here $\psi_{\sigma}^{\dag}(\vec r)$ and $\psi_{\sigma}(\vec r)$
are the creation and annihilation field operators for the two
fermionic species ($\sigma=\uparrow , \downarrow$). $V_{\sigma}
(\vec r)$ are the trapping potentials for the two species,
$V_{\Omega}(\vec r)$ is the optical potential created by the above
introduced two {\it co}-propagating laser fields with
Gaussian-Laguerre transverse profiles, $\mu_{\sigma}$ the chemical
potentials, $M$ the mass of the particles and the interaction
parameter $g=\frac{4\pi a_s \hbar^2}{M}$ with $a_s$ the scattering
length. Following a standard procedure \cite{jaksch} we get the
Hubbard Hamiltonian
\begin{eqnarray}
H=-\mu \sum_{m\sigma} c^{\dag}_{m\sigma} c_{m\sigma}
+\sum_{m \neq m',\sigma}\left(v_{mm'}c^{\dag}_{m\sigma} c_{m\sigma}
c_{m'\sigma}^{\dag} c_{m'\sigma}
- t_{mm'}c^{\dag}_{m} c_{m'}\right)
\end{eqnarray}
where the $v_{mm'}$ and $t_{mm'}$ determine atomic two-body
interactions and intersite hopping, respectively. We then reduce
the Hubbard model to the anisotropic Heisenberg model
\cite{anderson, bipolarons}. Let us now introduce fermionic pairs
formed by "spin-up" and "spin-down" particles via new $b_m$
operators
\begin{equation}
b_m\equiv c_{m\uparrow}c_{m\downarrow}
\end{equation}
The Hamiltonian turns out to become
\begin{eqnarray}
H=-\mu \sum_{m} b^{\dag}_{m} b_{m} +\sum_{m \neq
m'}\left(v_{mm'}b^{\dag}_{m} b_{m} b_{m'}^{\dag} b_{m'} -
t_{mm'}b^{\dag}_{m} b_{m'}\right)
\end{eqnarray}
Then using pseudo-spin representation for the $b_m$ operators
\begin{eqnarray}
b_{m}^{\dag}=S_{m}^{x}-iS_{m}^{y} \nonumber \\
b_{m}=S_{m}^{x}+iS_{m}^{y} \nonumber \\
b_{m}^{\dag}b_{m}=\frac{1}{2}-S_{m}^{z}
\end{eqnarray}
we arrive at the following anisotropic Heisenberg Hamiltonian
\begin{eqnarray}
H=\mu \sum_{m} S_{m}^{z}
+\sum_{m \neq m'}\left[v_{mm'}S^{z}_{m} S_{m'}^{z}
- t_{mm'}\left(S_{m}^{x} S_{m'}^{x}+S_{m}^{y} S_{m'}^{y}\right)\right]
\end{eqnarray}

{\it Bifermionic condensates and formation of ferro- and
antiferro-magnetic type crystalline structures}. The anisotropic
Heisenberg Hamiltonian was thoroughly studied in
\cite{bipolarons}. Let us define
$n_{b}=\left<b^{\dag}_{m}b_{m}\right>$ for bifermions, and $n_{a}$
for initially loaded atoms via
\begin{eqnarray}
\frac{1}{N}\sum_{m} \left< S^{z}_{m} \right> = \frac{1-n_{a}}{2}
\end{eqnarray}
We take into account only nearest-neighbor interactions having
$v_{m,m+1}=v$ and $t_{m,m+1}=t$. The model then possesses two
solutions. The first one is a uniform "ferromagnetic" state with
the number of bifermions $n_{b}=n_{a}/2$ uniformly distributed
along the ring and with the ground energy
$E=-\frac{1}{4}N\left[t+(t+v)(1-n_{a})^{2}\right]$. The second one
is the "anti-ferromagnetic" structure with the following solutions
for the number of bifermions for the two sublattices
\begin{eqnarray}
n^{(1)}_{b} =
\frac{1}{2}\left(n_{a}+\left[1+\left(1-n_{a}\right)^{2}-2v\left(1-n_{a}\right)/\sqrt{v^{2}-t^{2}}\right]\right) \nonumber \\
n^{(2)}_{b} =
\frac{1}{2}\left(n_{a}-\left[1+\left(1-n_{a}\right)^{2}-2v\left(1-n_{a}\right)/\sqrt{v^{2}-t^{2}}\right]\right)
\end{eqnarray}
and the ground state energy $E=-\frac{1}{4}Nv$. From the above
equations it follows that the antiferromagnetic structure forms
only for high enough atomic densities \cite{bipolarons}
\begin{eqnarray}
n_{a} \geq n_{c} = 1 -
\left[\left(v-t\right)/\left(v+t\right)\right]^{1/2}
\end{eqnarray}
Let us consider the number of bifermions vs the number of loaded
fermions. One clearly sees that the ground state of the system is
either uniform crystalline ordered phase or an antiferromagnetic
type structure with two sublattices containing unequal number of
biparticles $n^{(1)}_{b} \neq n^{(2)}_{b}$. The phase diagram of
the model anisotropic Heisenberg Hamiltonian was studied in detail
in \cite{T-n}. Their results show existence of four different
phases. Three of them below a critical temperature and two phases
above the temperature. Our calculation confirmed the critical
temperature obtained in \cite{bipolarons} that is
\begin{eqnarray}
T_{c} = \frac{
\left(1-n_{a}\right)t}{\ln\left[\left(2-n_{a}\right)/n_{a}\right]}
\end{eqnarray}
The critical temperature dependence on the number of loaded
particles is discussed in details in \cite{bipolarons}.

{\it Discussion}. The presence of two bifermionic condensate
phases as well as one non-condensate phase below the critical
temperature clearly indicates the qualitative difference between
the obtained superfluidity and a BCS type model. The model
considered in the present paper demonstrates existence of two
different non-condensate phases above the critical temperature
that also shows a drastic difference of the considered system from
the conventional ones. Another important property of the
considered model is the obtained possibility to form both
ferromagnetic and anti-ferromagnetic type crystalline structures.
They would present quantum crystals which can have clusters formed
by the model pseudospins representing the bifermions. It would be
very interesting to consider a possibility of using this system
for quantum computation via the pseudospin clusters in the manner
presented in \cite{spin_clusters}. Toroidal bifermionic
condensates may also attract an interest of the BEC community who
investigated superfluid toroidal currents of ultracold bosons
\cite{torus}.

{\it Acknowledgements}. The author is grateful to N.S. Ananikian
and K.V. Kheruntsyan for many valuable discussions. Different
parts of this work were discussed with E. Cornell, V. Gritsenko,
A. Ishkhanyan, V. Ohanyan, M. Roger, and A. Shames, whom I thank.

\end{document}